# OPINION SPAM RECOGNITION METHOD FOR ONLINE REVIEWS USING ONTOLOGICAL FEATURES

NGUYEN HOANG LONG,
PHAM HOANG TRONG NGHIA, AND NGO MINH VUONG

**ABSTRACT**

*Nowadays, there are a lot of people using social media opinions to make their decision on buying products or services. Opinion spam detection is a hard problem because fake reviews can be made by organizations as well as individuals for different purposes. They write fake reviews to mislead readers or automated detection system by promoting or demoting target products to promote them or to damage their reputations. In this paper, we propose a new approach using knowledge-based Ontology to detect opinion spam with high accuracy (higher than 75%).*

**TÓM TẮT**

***Phương pháp nhận diện nhận xét rác cho các ý kiến trực tuyến sử dụng các đặc điểm ontology***

*Ngày nay, rất nhiều người tham khảo ý kiến trên các phương tiện truyền thông nhằm quyết định mua các sản phẩm hoặc dịch vụ nào đó. Việc phát hiện ý kiến rác là vấn đề khó bởi vì các nhận xét lừa đảo có thể được viết ra bởi các tổ chức cũng như cá nhân với nhiều mục đích khác nhau. Họ viết các nhận xét lừa đảo này nhằm mục đích đánh lừa người đọc hoặc hệ thống nhận diện tự động để đề cao sản phẩm của họ hoặc đánh giá thấp các sản phẩm đối thủ. Trong công trình này, chúng tôi đề xuất một hướng tiếp cận khác, đó là sử dụng Ontology làm cơ sở tri thức để giải quyết bài toán nhận diện nhận xét rác, với độ chính xác đạt được trên 75%.*

## 1. Introduction

Most e-commerce websites now allow users to leave reviews of the products that they have used or traded directly on these websites. Reviews of a product are defined as the individual assessment of the product or service 1. Reviews must contain information about quality, or characteristics of the product. The reviews have become a good resource for decision making. In recent years, along with web spam 19, 22, email spam 23, 10 and blog spam 20, 18, review spam detection has attracted attention from research community 11, 14.

Reviews on products are very important for both sellers and buyers in purchasing online. Customers who use the service from e-commerce websites will reference information from other customers through these reviews and make the best decision when they intend to buy a product. Suppliers also base on reviews to learn about customer opinions and customer demands in order to analyses and come up with strategies necessary for their business.

A review consists of 3 main components:

1. Category: type of products reviewed, such as: phone, camera and tablet.
2. Name: series of product being reviewed. For example, products of phone type will have names such as: iPhone, Samsung Galaxy and LG.
3. Content: text that contains the entire opinions of users.

Jindal and Liu 7 have classified spams into three main types, which are: non-reviewed, brand-only review and untruthful review.

1. Non-reviewed review consists of two main types: First, comments that do not contain opinion meaning, or they cannot express any idea, or views of the users of the product reviewed. The second form is advertisement. This type often shows advertisement for business target.
2. Brand-only review is a type where contents are not direct evaluation of specific products but assess of the company or suppliers of those products.
3. Untruthful review, also known by several common names, such as: fake review or deceptive review. Comments of this type are often deliberately either positive or negative reviews about a certain product to deceive users.

In addition to the three types of comment spam, we propose to add a fourth review: off-topic review. The content of the off-topic review is not related to the reviewed product. For example, although iPhone 4S is the product being assessed, its reviews have content mentioning Samsung Galaxy S4. This kind of review should be removed because the readers only need the necessary information.

## 2. Related works

### *2.1. Content-based spam detection*

To classify comment spam, Ott, et al. 17 have used a classification model based on machine learning using Naïve Bayes and Support Vector Machine. Specifically, there are 3 approaches for this problem: genre identification, psycholinguistic and text categorization. The feature sets that are used for training the classification models include POS, LIWC, UNIGRAM, BIGRAM+, and TRIGRAM+. In addition, the authors also combined two feature sets of LIWC and BIGRAM+. Experiments show that this combination approach achieves higher accuracy than using a single feature set.

While Ott, et al. 17 have used the complex technique of natural language processing and focused on the psychological field of the reviews, the authors of 7 have proposed a simpler strategy which is to use sets of duplicate reviews based on three popular models: logistic regression, SVM, and Naïve Bayes.

Another feature of opinion spam which is recently researched is the utility of a review. A method proposed and studied in Zhang and Varadarajan 21 is utility scoring.

Useful review is a review that is reliable and contains useful information for the reader. An example of useless reviews is neutral review, e.g. a review that does not show opinion clearly and readers will be so confused when making a decision.

### *2.2. Behavior-based spam detection*

There are many types of abnormal behavior. Experiments show that reviews written by these people are likely to be spam reviews, mentioned in 8 and 11. The first method is to seek for unusual patterns using unexpected law 8. The approach of this study is to identify the unusual patterns in the review, the review followed with abnormal behavior of the reviewer. With this approach, domain independent technique will be used to build the unexpected law. The data is a set of basic attributes: $A = \{A_1, \ldots A_n\}$ and set of classification attributes: $C = \{c_1, \ldots, c_m\}$, C includes m discrete values. The law will be expressed as: $X \rightarrow c_i$, where X is the set of conditions from the attributes of A and $c_i$ is a class in C. With each law: the conditional probability $Pr(c_i|X)$ (also called reliability) and the probability $Pr(X, c_i)$.

Another method was introduced in 11 is scoring behavior and detecting spammers. Data set is a collection of user reviews for the product. Products collected from website amazon.com. Based on spam pattern extracted from data set, the study identified the following unusual behaviors: (1) targeting products; (2) targeting product groups; (3) general deviation and (4) early deviation. Finally, evaluation function will be built to score each user based on abnormal behaviors mentioned above. Final spam score will be combined by spam scores of four behaviors.

The above works studied spam review by analyzing one aspect of the review, they are content of the review and behavior of the reviewer, it is called single-view algorithm. A method is proposed for optimizing learning algorithm is two-view co-training algorithm mentioned in 6. Experimental work has proved that a spammer's review has a probability of 85% to be a spam review. Thus, proving whether the author of the review is a spammer or not is the main task of this classification model. Result after using the algorithm is a classifier that has the ability to identify spam review based on the content and the probability of whether the author is spammer or not.

Besides, in 14, the authors have proposed a collaborative setting method to discover fake reviewer groups. The method finds a set of candidate groups by item set mining before using some behavioral models. The experiment results showed that the proposed relation-based model significantly outperformed the state-of-the-art supervised classification model.

In 2, the authors have exploited the business nature of reviews to identify review spammers. In addition, the authors also build a network of reviewers appearing in different bursts and exploit the Loopy Belief Propagation method to infer whether a reviewer is a spammer or not in the graph.

The method in Mukherjee, et al. 12 have used Bayesian model to exploit observed reviewing behaviors to detect fake reviewers. Bayesian inference can character-ize facilities of various spamming activities by using the estimated latent population distributions.

### 2.3. Other studies

Besides, there are other spam studies such as: group spam or web spam. In 13, authors have given a method to detect spam in groups, and conducted step by step as follows: first, mining frequent patterns to find candidate group; second, authenticate candidate group by using the criteria of unusual behavior and finally ranking the candidates. Results returned from the ranking function will then be learned by the SVM learning and conducting final classification for the group of candidates.

In the study of spam, web spam has been researched for a long time and had a lot of practical applications. Web spam is defined as a website containing spam content or unexpected content for readers that disturbs them when surfing the web. Most spam sites try to take advantage of SEO techniques to increase their ranking on search engines, then gain more readers and achieve advertising purposes or vandalism. Ntoulas, et al. 16have proposed few approaches to classify web spam, as well as some experiences are designed to optimize the problem. The research focused on web spam classification model using content analysis method, so that content-based experience was used for training model, including: the number of words contained in the web page, the number of words contained in the web page title, the average length of words, number of anchor text, the compress ratio, the ability to compress content. In 4, the authors have created a dataset including 400 fake reviews and 400 true reviews. After that, they use a method combining human-based assessment and machine-based assessment.

## 3. Knowledge base

### 3.1. Ontology and OWL

In computer science, Ontology is defined as a data model used to represent a concept about a certain area and relationships between them 3, 9. Ontology model includes a vocabulary data used to describe the concept of a particular field. In addition, Ontology also includes the meaning of each word in the vocabulary. Ontology is usually used in the field of artificial intelligence, natural language processing, semantic web, information system, etc. It is a useful tool to conceptualize the knowledge base of a particular field in a database format that computer can understand 15. Most ontology consists of 4 main components: objects (instances), classes (concepts), attributes and relations.

OWL (The Web Ontology Language) 5 is a language for publishing and sharing data over the Internet via the data model called the "Ontology". OWL builds on RDF platform. OWL is a markup language like XML which is almost used to describe entities, classes, attributes and relationships between them, but is wider than the RDF Schema. All these factors, the nature of RDF and RDF Schema can also be used to generate an OWL documents. OWL provides a data model and a simple syntax so that independent systems can share and use it. In addition, it is designed not only for people using it but also for the computer system so that it can understand and exploit information. The main purpose of OWL is to create a standard platform to manage resources on the Web.

### *3.2. POS tagging and grammar parser*

POS tagging is the identification of all kinds of words in a context. POS tagging is a very important operation and required for all systems of natural language processing. It is the first step in the analysis of multiple parsing. About applications, POS tagging is useful in many fields of information retrieval, voice synthesis, research compiled dictionaries, terminology mining and many other applications. The following example illustrates the activity of POS tagging:

*My dog also likes eating sausage*

A sentence with such content, after being processed will result in:

*My/PRP$ dog/NN also/RB likes/VBZ eating/VBG sausage/NN*

Parsing is defined as the process of analyzing a text and gives a description of the grammatical structure of the components (the sentence, the terms, phrases and words) of the documents. The model is based on a set of operational constraints on the syntax of a language, such as: S $\rightarrow$ NP VP. First, with a text that is inserted, the text will be labeled and after labeling, each word is defined morphological characteristics. Then a process of checking syntax and combining of words will be conducted for the input, based on the syntax rules for removing cases of irregularity and gradually build up syntactic structures (parse tree) of the sentence. Here, results returned after conducting parsing the above sentence are displayed in a parser tree.

*(ROOT*
*(S*
*(NP (PRP$ My) (NN dog))*
*(VP*
*(ADVP (RB also))*
*(VBZ likes)*
*(NP (JJ eating) (NN sausage)))))*

## 4. Proposed model

### *4.1. Ontology model*

Ontology is not able to cover all aspects of a study field. With the specific objective of identifying spam review, extracted entities also focused on components or properties of reviewed products. A number of related entities can be ignored to avoid ambiguity for Ontology.

An ontology is impossible to cover all meaning aspects of a field, so that the specific objectives is used for identifying spam review, extracted entities also focused on components or properties of product. A number of related entities can be ignored to avoid ambiguity for Ontology later. Thus, the entity after the statistics will be collected and distributed to the class groups based on their common characteristics.

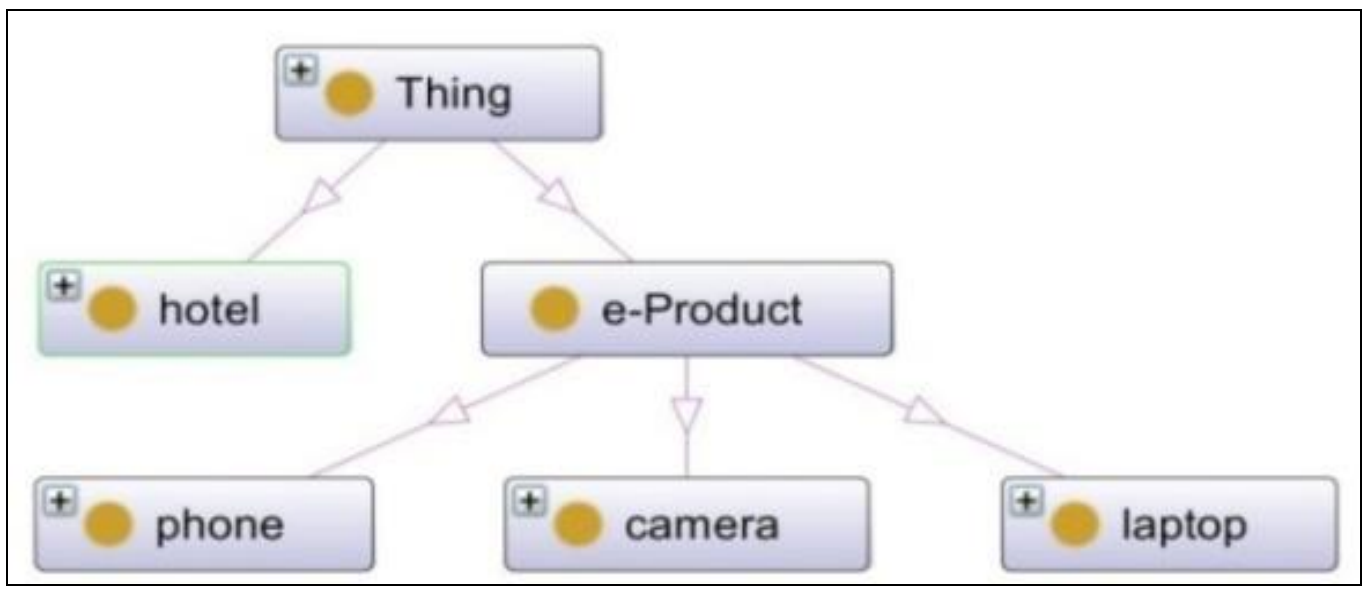

***Figure 1 .****Ontology model*

Figure 1 presents the classes containing information products. Most general class *Thing* is divided into two subclasses *e-Product* and *hotel* for two selected products. For class *e-Product*, we selected three most popular *e-Product* includes: *phone, laptop*, and *camera*. Based on the statistical entity from the data set, each *e-Product* will be divided into four main classes:

1. Component/Feature: contains objects describing the composition, hardware or software of products.

2. Style: contains objects describing the design, product design.

3. Origin: contains objects describing the origin, brand of the product. This is an important class in the ontology, which supports the brand-only and off-topic detection algorithm.

4. PopularName: contains the name of the popular products of this product. For example, the phone will have the class name of popular products such as: iPhone, GalaxyS3, Onex and GalaxyNote.

Depending on the type of product, each class is further broken down in order to better describe the meaning of each class. Class *component* can be divided into *software* and *hardware* or class *style* can be broken down into *color* and *design*. Table 1 presents a number of subclasses and entities belongs to the five most popular classes.

***Table 1.*** *Statistic table of classes and entities in the Ontology model*

| | e-Product | | | hotel | Total |
|---|---|---|---|---|---|
| | phone | camera | laptop | | |
| Class | 26 | 7 | 27 | 3 | 63 |
| Entities | 211 | 95 | 181 | 81 | 568 |

### *4.2. Preprocessing module*

Preprocessing module is responsible for analyzing content and title of review and producing the necessary data for the classification model. Preprocessing work is divided into four sub modules as the diagram in Figure 2.

1. Entities building module: Consider product type as an input, this module is responsible for retrieving the knowledge base from Ontology and extract all entities from the corresponding branch of this product.

2. Normalizing module: The content of the review is the most important input of the model. Therefore, before proceeding with the other processing steps, normalizing needs to be done to create standard data sources and avoid errors analyzing.

3. Word splitting and grammar parser module: There are many different approaches to split words from a text. Within the scope of this study, we have chosen the method n-gram models of unigram, combined with the POS tagging model. The POS tagging tools of Stanford University (Stanford POS Tagger) has a set of fairly large databases and has been widely used in the study of language processing. With the advantages of high accuracy and processing performance, we have chosen to use this tool for word splitting module.

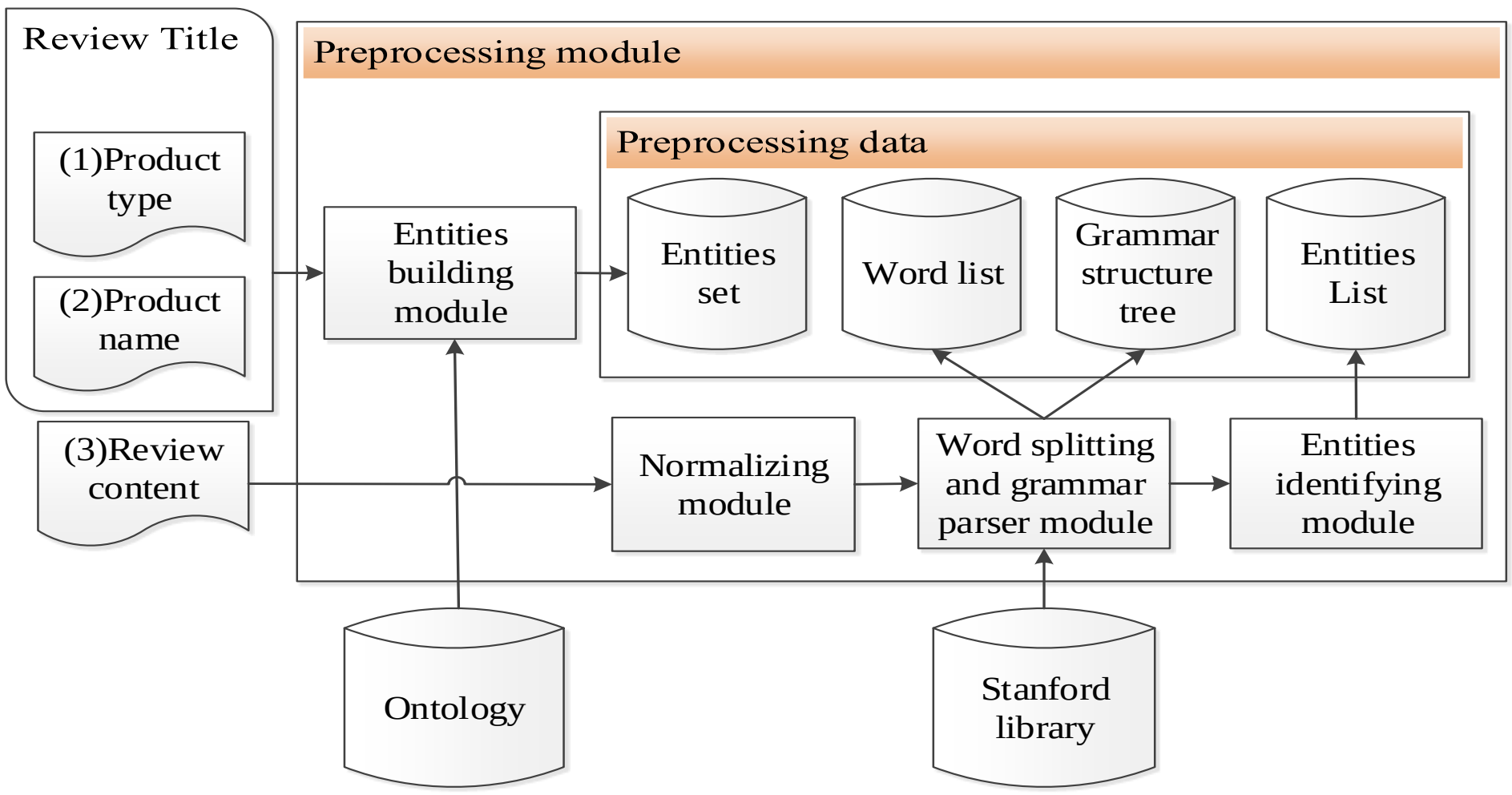

***Figure 2.*** *Workflow of pre-processing module*

For example: given the following review: *Samsung is the world's leader in screens(Sony is a part of them) the next Samsung is obviously going to have a better screen, they are already in the process of making one that is unmatched in the mobile world....apple does not make their own hardware*

The preprocessing module will process and produce a collection of data:

i: Entities list: entities contained in the review: Samsung, world, screen, Sony, mobile, apple, hardware.

ii: Entities set: based on the product type (mobile) and ontology modal, the module will also produce an entities set retrieved from the ontology tree branch of mobile. This set may or may not contain the entire entities list above.

4. Entities identifying module: The entity is a crucial component in the spam recognizing system, being the base knowledge for searching process and matching

Ontology. A sentence can contain just one entity or multiple entities, or even without any entities. To support the algorithm to identify spam review, preprocessing modules will perform and recognize this kind of entity and save the entity identified in the data preprocessing. The entity here is not just the named entities but the entities in general in the sense that the researchers defined. With the aim of using these entities to find the product knowledge contained in the reviews, we define entity as the word meaning, which brings specific knowledge in the reviews. According to this definition, adjective and noun are two words of which we have chosen to filter into the desired entity.

### *4.3. Opinion spam detection module*

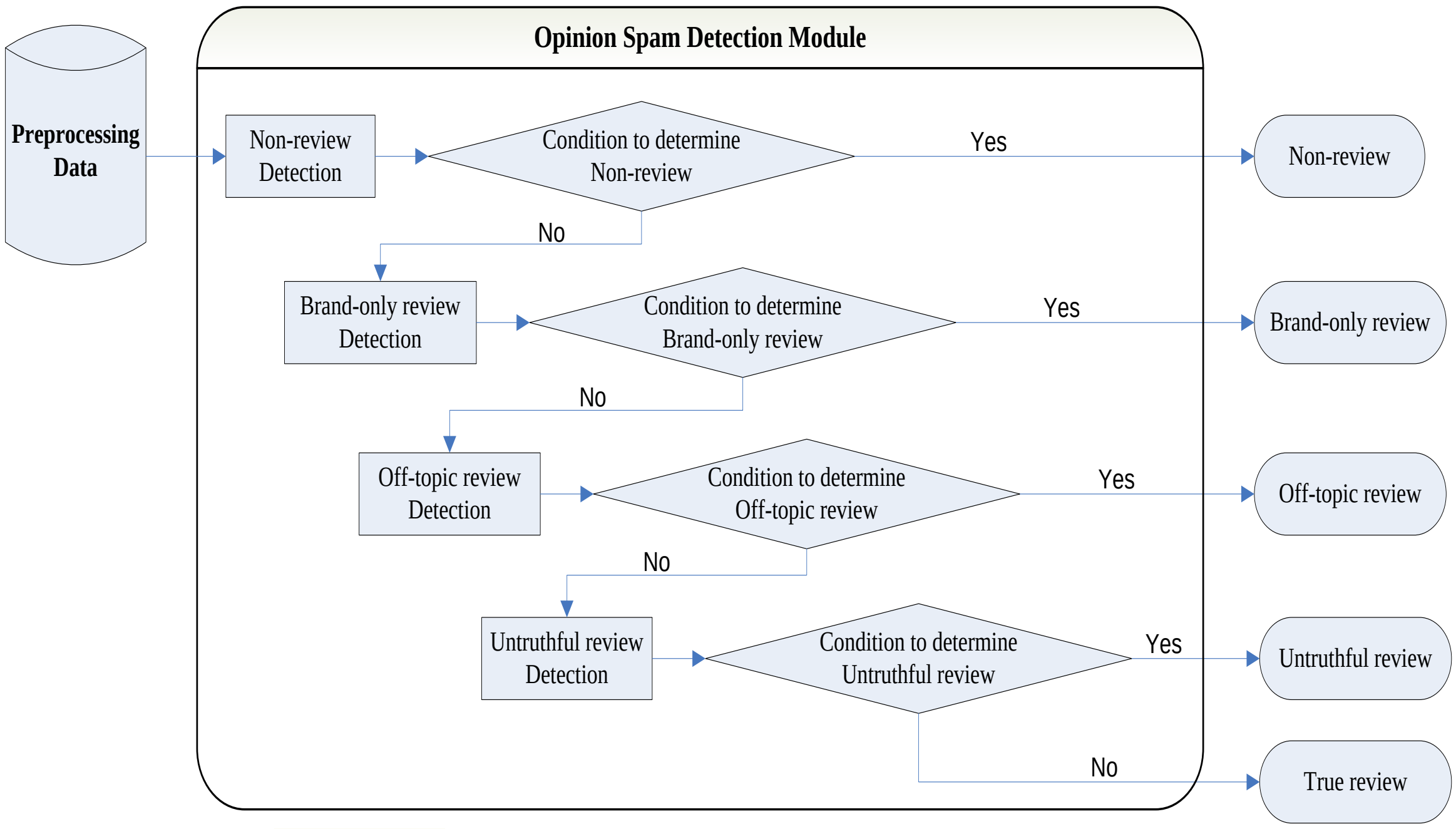

***Figure 3.*** *Workflow of opinion spam detection module*

We also develop an Opinion Spam Detection Module to handler and process the preprocessing data above, then produce the final result of detection.

Opinion spam detection module is responsible for clustering reviews into fake reviews and true reviews. In fake reviews, there are four sub-types: non-review, band-only review, off-topic review and untruthful review. Figure 3 presents how works of opinion spam detection module.

#### *4.3.1. Non-review detection*

1. Finding unusual pattern: Based on experiments, we found that non-review contains a lot of unusual patterns. Unusual patterns are defined ad advertisements, links to other sites, email addresses, phone numbers, and price. The probability of a non-review is increasing when there are more unusual patterns appearing in the review.

2. Opinion word ratio: A true review needs to provide readers with knowledge about the product. Therefore, to achieve a condition to be a review, the content of the statement must contain at least a number of opinion words, compared with the total number of words in a review. Non-review contains very little opinion words, or even contains no opinion words.

3. Ontology word ratio: As described in Section 4.1, Ontology is a base knowledge that contains attributes and knowledge about product. Review containing the entity that cannot achieve this knowledge will be classified as non-review.

4. Sentence ratio: According to our survey, a number of non-reviews can be written with standard words, no random characters and no unusual patterns that we mentioned. However, the combination of these words are completely meaningless; in other words, the syntax is wrong. Grammar parser will be used to determine whether the structure of a text is a sentence or not. Thus, based on the sentence list, mapped with their syntax, this module will calculate a percentage of meaningful structures to use as a condition for classifying reviews.

Example: Non-review

<3<3<3<3<3<3<3<3<3<3<3<3<3<3<3<3<3<3<3<3<3<3<3<3<3<3<3<3<3<3 Great!!!!!!!!!!!!!!!!!!!!!!!!!!!!!!!!!!!!!!!!!!!!!!!!!!!!!!!!!!!!!!!!!!!!!!!!!!!!!

*4.3.2. Brand-only review detection*

As described above, in the ontology that we build, each product branch contains Origin class that contains entities describing its product name, manufacturer name, distributor, or country of manufacturing. For example, the class origin of the phone product branch may include the following entities: Apple, Samsung, Amazon, HTC, LG, company, seller, Korea, America.

The main task of branch-only review modules is to count entities in the review that belong to class origin and output percentage calculations.

Example: Brand-only review

Samsung is the world's leader in screens(Sony is a part of them) the next Samsung is obviously going to have a better screen, they are already in the process of making one that is unmatched in the mobile world....apple does not make their own hardware.

*4.3.3. Off-topic review detection*

To identify off-topic reviews, we divided this review type into two sub types as follows:

1. Off-type: review talks about another type of product.

2. Off-brand: review talks about the same type of product, but another name or brand of product.

Consider product type as input, pre-processing module will identify the branch of the ontology contains the knowledge base of this product. The task of the module is to match entities in the review with corresponding entities set, and to calculate the condi-tional rate of classification. The purpose of this matching is to determine whether the composition of the products reviewed corresponds to the actual composition of it or not. The meaning of this ratio is described as follows: if the entities that do not belong to product brand appear with a great number, the review is talking too much about another product, or another brand of product.

Above conditions is not enough to identify off-brand review. Another condition is added: The percentage of product name entities does not coincide with the name of the product reviewed. Similarly, this ratio tells us off-brand degree of a review.

Example: Off-topic review

And the mac is also a better built computer. I was talking to a guy the other day and his MacBook pro was stolen and then people who stole it smashed it and ran it over and apple employee found it and returned it to him and the data was recovered but the screen was ruined.

*4.3.4. Untruthful review detection*

1. Opinion polarizing ration: If a review is only to praise or to criticize a product and does not have much opinion, these reviews are totally suspicious. We undertook to build module that calculate the attitude and opinion word by the review. The outcome of this module is the statistics on the number of expressed positive attitudes and negative attitudes in the reviews and polarizing opinion ration.

2. Duplicate name ratio: With the usual reviews reviewers mention the product name only in the first sentence of the text to introduce and then, in the next sentence, they will go straight to product characteristics, strength and weakness. Product names can then be referred to pronoun. Spammer often repeat product name for advertising purposes in order to attract the reader. Duplicate name ration will be calculated based on this experience.

3. Capitalize name ratio: One of the ways for manufacturers as well as those who are hired to write a review for the product impression is deliberately capitalized name brands. With the usual review, the reviewer simply lowercases names or products, some capitalize only the first letter of the name brands. We use this experience to detect untruthful review.

4. Extreme word ratio: Reviewers do not tend to write as many positive or negative words as possible for their target products than those who are less likely to have experience of the product, even never use the product. As a psychologist who always wanted to read the comments that people write their reviews were accidentally use too many words to say so that they themselves did not know. For English, through the reference material and the samples were tested to comment on the episode, we see that from talking so here is the adverb, which is all positive, or all targets extreme. When added to a sentence adverb, the sentence stress levels will raise so much.

Example: Untruthful review

I stayed at the Sheraton Chicago Hotel for two nights and I must say the service they rendered was quite impressive. They had very attentive and friendly staff members. The room that I stayed in was spacious for me and my husband. Their hotel restaurant served the most delicious steaks I have tasted, I ordered a classic fillet mignon and it was cooked to perfection. I would definitely stay at this hotel again if ever I come back to Chicago and would absolutely highly recommend this to my friends and family.

### *4.4. Threshold:*

The threshold to determine a review is spam or not will be learnt by processing the dataset. With the dataset, our system will change the threshold; calculate the measurement and record best result in order to have the most appropriate threshold value for the system.

## 5. Experimental result

### *5.1. System testing*

To test the system, we choose the two following products to build the data set:

1. E-product: we choose three popular products (camera, laptop and mobile) that attract many users' reviews. These reviews were selected from popular e-commerce websites such as amazon.com, ebay.com and partial data set used in Jindal and Liu 7.

2. Hotel: we selected and extracted reviews from the available data set that is used in Ott, et al. 17. These reviews' quality were evaluated and re-examined carefully by the authors through different methods which include mechanical inspection and manual inspection.

Basing on the above data sources, we conducted to build two data sets. Each data set includes 800 review which be divided into two parts presented in Table 2. We use following measure tool to evaluate performance of the model: Precision (P), Recall (R) and F-measure (F). In pattern recognition and information retrieval with binary classification, precision (also called positive predictive value) is the fraction of retrieved instances that are relevant, while recall (also known as sensitivity) is the fraction of relevant instances that are retrieved. Both precision and recall are therefore based on an understanding and measure of relevance. In simple terms, high precision means that an algorithm returned substantially more relevant results than irrelevant, while high recall means that an algorithm returned most of the relevant results. Usually, precision and recall scores are not discussed in isolation. Instead, either values for one measure are compared for a fixed level at the other measure or both are combined into a single measure. Examples for measures that are a combination of precision and recall are the F-measure (the weighted harmonic mean of precision and recall)

| Classify / Expectation | Spam | Non-spam |
| --- | --- | --- |
| Spam | tp | Fp |
| Non-spam | fn | tn |

$$P = \frac{tp}{tp + fp}$$

$$R = \frac{tp}{tp + fn}$$

$$F = 2 \times \frac{P * R}{P + R}$$

***Table 2.*** *Statistic of reviews for each type in the dataset*

| | Non-review | Brand-only review | Off-topic review | Untruthful review | Truthful review |
|---|---|---|---|---|---|
| **DataSet 1** | 100 | 100 | 100 | 100 | 400 |
| **DataSet 2** | 100 | 100 | 100 | 100 | 400 |

The chart in Figure 4 shows that evaluative performance of the second data set slightly lower than the first one. The majority of the first data set is reviews used in Ott, et al. [17] is the main reason, which were filtered and re-examined very carefully through several steps that include mechanical inspection and manual inspection by the authors. Thus, it can be concluded that noise and the accuracy of the data set also partially affected the performance of system, especially for untruthful reviews, so a really standard data set to ensure enough accuracy when evaluating performance is very necessary.

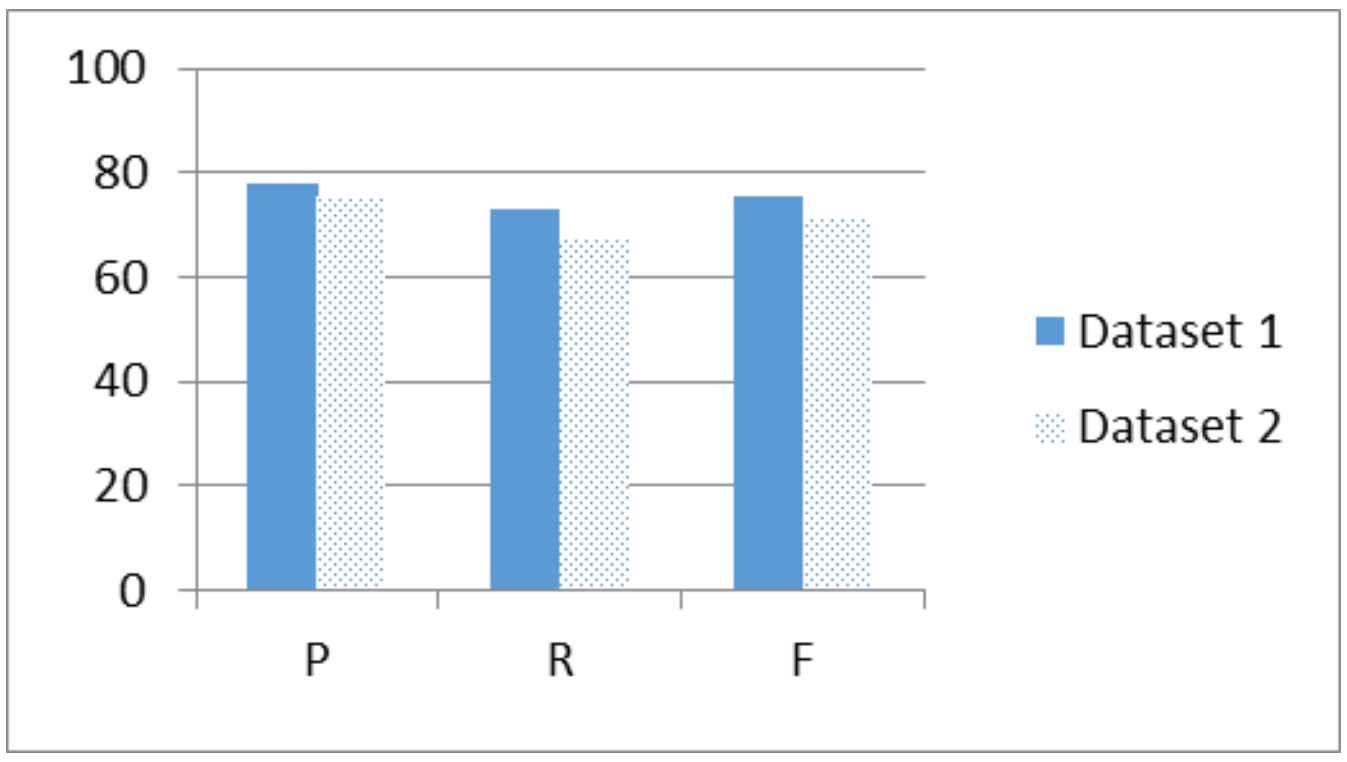

***Figure 4.*** *The evaluation result of system on 2 data set*

### *5.2. Module testing*

Besides evaluating performance of the system, we also evaluated the performance of each module which corresponding to each type of spam review. To evaluate the performance of each module, we divides the data set which consists of 800 comments into four parts presented in Table 3, each part has 200 reviews in which 100 reviews corre-spond to the module that need to be evaluated and 100 truthful comments are taken randomly.

***Table 3.*** *Divison of sub data set*

| | **Subset 1** | **Subset 2** | **Subset 3** | **Subset 4** |
|---|---|---|---|---|
| **Spam review** | 100 non-review | 100 brand-only | 100 off-topic | 100 untruthful |
| **Non-spam review** | 100 truthful | 100 truthful | 100 truthful | 100 truthful |
| **Sum** | 200 review | 200 review | 200 review | 200 review |

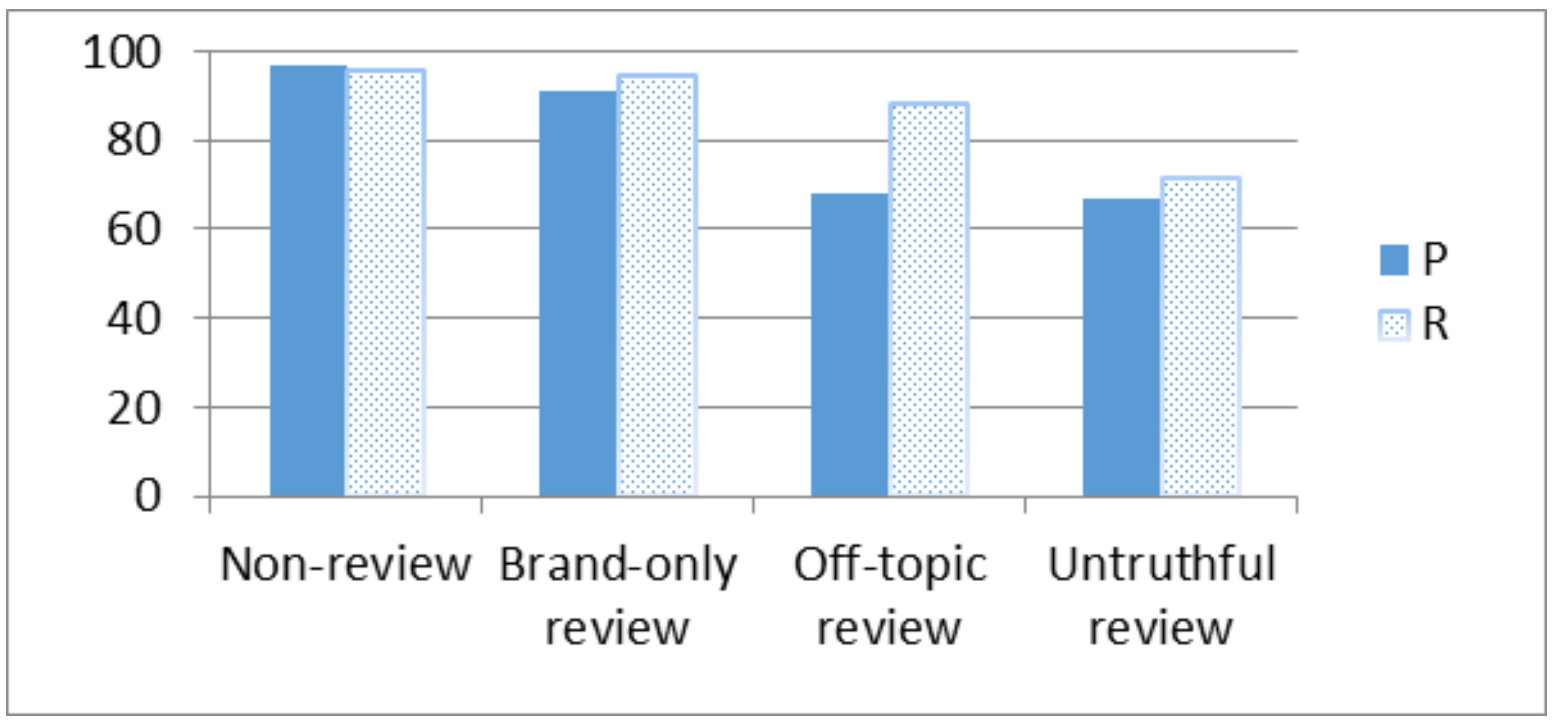

***Figure 5.*** *The evaluation result of system on 4 type of spam review*

The chart in Figure 5 compares the performance of each module. It shows that the non-review detection module has the best performance compared to other modules (over 95%). By adding unusual patterns to the dictionary, performance of the non-review detection module was improved significantly. Conversely, the brand-only detection module has the lower performance compared to the rest. As we have described in section 4, there are a lot of experiences to identify whether a review is spam or not. However, we can't apply all experiences to the model because of the proposed model using Ontology and recognizing spam review based on their content. This is also a goal for us to improve the system in the future.

## 6. Conclusions

This research focused on analyzing spam review based on their content, which combines with Ontology model as the main model in designing algorithm to identify these remarks. We divide junk remarks into four types: non-review, brand-only review, off-topic review and untruthful review. In which, we re-use three kinds of spam review from previous studies and add the off-topic review. Two data sets with 800 reviews each were collected to check performance of the system that we have built. These two sets are classified and labeled corresponding to four types of spam review and truthful review. With two data sets, the system give out a quite good result in classification which shows that performance of the system reached over 75% (P). With each identifying module, non-review identifying module produces the result of classification which achieved over 90%, while the three rests have lower performance. In our future work, we will improve our system by combining the proposed method with structure mining of reviews, and also look into spam in other kinds of media, e.g., social networks. Moreover, we also have some idea to build a system which can recognize the Vietnamese spam reviews, also based on the advantages of Ontology, for the reviews on e-commerce section.

## REFERENCES

1. Duong, T. H. H., Dai, T. V. and Ngo, V. M. (2012), "Detecting Vietnamese Opinion Spam", *In Proceedings of Scientific Researches on the Information and Communication Technology in 2012 (ICTFIT'12)*, pp. 53-59.
2. Fei, G., et al. (2013), "Exploiting Burstiness in Reviews for Review Spammer Detection", *Proceedings of ICWSM.*
3. Fensel, D., Harmelen, V. F.and Horrocks I. (2001), "OIL: An Ontology Infrastructure for the Semantic Web", *IEEE Intelligent system*, 16(2), pp. 38-45.
4. Harris, C. G. (2012), "Detecting Deceptive Opinion Spam using Human Computation", *In Proceedings of the 4th Human Computation Workshop (HCOMP'12)*, pp. 87-93.
5. Hitzler, P., et al. (2012), "OWL 2 Web Ontology Language Primer (Second Edition)", *In World Wide Web Consortium*.
6. Huang, M., Yang, Y. and Zhu, X. (2011), "Learning to Identify Review Spam", *In Proceedings of IJCAI*, pp. 2488-2493.
7. Jindal, N. and Liu, B. (2008), "Opinion Spam and Analysis", *In Proceedings of the international conference on Web search and web data mining (ACM'08)*, pp. 219–230.
8. Jindal, N., Liu, B. and Lim, E. P. (2010), "Finding Unusual Review Patterns Using Unexpected Rules", *In Proceedings of the 19th ACM CIKM*, pp. 1549-1552.
9. Junwu, Z., Bin, L., Fei, W. and Sicheng, W. (2010), "Mobile Ontology", *In Proceedings of International Journal of Digital Content Technology and its Applications (JDCTA),* 4(5), pp. 46-54.
10. Li, J., Deng, X. and Yao, Y. (2013), "Multistage Email Spam Filtering Based on Three-Way Decisions", *In Proceedings of 8th International Conference on Rough Sets and Knowledge Technology, LNCS, Springer*, pp. 313-324.
11. Lim, E.P., Nguyen, V.A., Jindal, N., Liu, B. and Lauw, H.W. (2010), "Detecting Product Review Spammers using Rating Behaviors", *In Proceedings of the 19th ACMCIKM.*
12. Mukherjee, A., et al. (2013), "Spotting Opinion Spammers using Behavioral Footprints", *In Proceedings of 19th SIGKDD.*

13. Mukherjee, A., Liu, B., Wang, J., Glance, N. and Jindal, N. (2011), "Detecting Group Review Spam", *In Proceedings of the 20th WWW*, pp. 93-94.

14. Mukherjee, A., Liu, B. and Glance, N. (2012), "Spotting Fake Reviewer Groups in Consumer Reviews", *In Proceedings of the 21st WWW*, pp. 191-200.

15. Ngo, V. M. and Cao, T.H. (2011), "Discovering Latent Concepts and Exploiting Ontological Features for Semantic Text Search", *In Proceedings of the 5th IJCNLP*.

16. Ntoulas, A., Najork, M., Manasse M. and Fetterly, D. (2006), "Detecting Spam Web Pages through Content Analysis", *In Proceedings of the 15th WWW*, pp. 83-92.

17. Ott, M., Choi, Y., Cardie, C. and Hancock, J. T. (2011), "Finding Deceptive Opinion Spam by any Stretch of the Imagination", *In Proceedings of the 49th ACL-HTL*.

18. Prieto, V. M., Álvarez, M. and Cacheda, F. (2013), SAAD, "AContent based Web Spam Analyzer and Detector", *In Journal of Systems and Software (JSS)*, 86(11), pp. 2906–2918.

19. Spirin, N. and Han, J. (2012), "Survey on Web Spam Detection: Principles and Algorithms", *In ACM SIGKDD*, 13(2), pp. 50-64.

20. Teraguchi, T., et al. (2012), "Detection Method of Blog Spam Based on Categorization and Time Series Information", *In Proceeings of 26th International Conference on Advanced Information Networking and Applications Workshops (WAINA)*, pp. 801-808.

21. Zhang, Z. and Varadarajan, B. (2006), "Utility Scoring of Product Reviews", *In Proceedings of the 15th ACM CIKM*.

22. Zhang, H., Jin, D. and Zhao, X.J. (2014), "Research on Information Technology with Detecting the Fraudulent Clicks Using Classification Method", *In Advanced Materials Research*, 859, pp. 586-590.

23. Zhou, B., Yao, Y. and Luo, J. (2013), "Cost-sensitive Three-way Email Spam Filtering", *In Journal of Intelligent Information Systems (JIIS)*, pp. 1-27.